\documentclass[aps,prl,reprint,superscriptaddress]{revtex4-2}
\usepackage{graphicx} 
\usepackage{dcolumn}  
\usepackage{bm}       
\usepackage{amsmath, amssymb, amsthm}  

\usepackage{verbatim}
\usepackage{xcolor}
\usepackage{tikz} 
\usepackage{braket}

\newcommand{\plaquette}{%
    \begin{tikzpicture}[baseline=-0.5ex, scale=0.25]
        \draw[thick] (0,0) -- (1,0);
        \draw[thick] (0,0) -- (0,1);
        \draw[thick] (0,1) -- (1,1);
        \draw[thick] (1,0) -- (1,1);

        \filldraw (0,0) circle (4pt);
        \filldraw (0,1) circle (4pt);
        \filldraw (1,0) circle (4pt);
        \filldraw (1,1) circle (4pt);
    \end{tikzpicture}%
}

\newcommand{\adjplaquettes}{%
    \begin{tikzpicture}[baseline=0.5ex, scale=0.2]
        \draw[thick, color = black] (0,0) -- (0,1.6);
        \draw[thick, color = black] (0.8,0) -- (0.8,1.6);
        \draw[thick, color = black] (0.0,0) -- (0.8,0);
        \draw[thick, color = black] (0,0.8) -- (0.8,0.8);
        \draw[thick, color = black] (0,1.6) -- (0.8,1.6);
        
        \filldraw[color = black] (0,0) circle (4pt);
        \filldraw[color = black] (0,0.8) circle (4pt);
        \filldraw[color = black] (0.8,0) circle (4pt);
        \filldraw[color = black] (0.8,0.8) circle (4pt);
        \filldraw[color = black] (0,1.6) circle (4pt);
        \filldraw[color = black] (0.8,1.6) circle (4pt);
    \end{tikzpicture}%
}

\newcommand{\vplaquette}{%
    \begin{tikzpicture}[baseline=0.2ex, scale=0.3]
        \draw[thick, color = black] (0,0) -- (0.75,0);
        \draw[thick, color = black] (0,.75) -- (0.75,0.75);
        \filldraw[color = black] (0,0) circle (4pt);
        \filldraw[color = black] (0,0.75) circle (4pt);
        \filldraw[color = black] (0.75,0) circle (4pt);
        \filldraw[color = black] (0.75,0.75) circle (4pt);
    \end{tikzpicture}%
}

\newcommand{\vplaquettered}{%
    \begin{tikzpicture}[baseline=0.2ex, scale=0.3]
        \draw[thick, color = red] (0,0) -- (0.75,0);
        \draw[thick, color = red] (0,.75) -- (0.75,0.75);
        \filldraw[color = red] (0,0) circle (4pt);
        \filldraw[color = red] (0,0.75) circle (4pt);
        \filldraw[color = red] (0.75,0) circle (4pt);
        \filldraw[color = red] (0.75,0.75) circle (4pt);
    \end{tikzpicture}%
}

\newcommand{\vplaquetteblue}{%
    \begin{tikzpicture}[baseline=0.2ex, scale=0.3]
        \draw[thick, color = blue] (0,0) -- (0.75,0);
        \draw[thick, color = blue] (0,.75) -- (0.75,0.75);
        \filldraw[color = blue] (0,0) circle (4pt);
        \filldraw[color = blue] (0,0.75) circle (4pt);
        \filldraw[color = blue] (0.75,0) circle (4pt);
        \filldraw[color = blue] (0.75,0.75) circle (4pt);
    \end{tikzpicture}%
}
\newcommand{\hplaquette}{%
    \begin{tikzpicture}[baseline=0.2ex, scale=0.3]
        \draw[thick, color = black] (0,0) -- (0,0.75);
        \draw[thick, color = black] (0.75,0) -- (0.75,0.75);

        \filldraw[color = black] (0,0) circle (4pt);
        \filldraw[color = black] (0,0.75) circle (4pt);
        \filldraw[color = black] (0.75,0) circle (4pt);
        \filldraw[color = black] (0.75,0.75) circle (4pt);
    \end{tikzpicture}%
}

\newcommand{\horizontaldimer}{%
    \begin{tikzpicture}[baseline=-0.5ex, scale=0.3]
        \draw[thick, color = black] (0,0) -- (0.75,0);
        
        \filldraw[color = black] (0,0) circle (4pt);
        \filldraw[color = black] (0.75,0) circle (4pt);
    \end{tikzpicture}%
}

\newcommand{\horizontaldimerred}{%
    \begin{tikzpicture}[baseline=-0.5ex, scale=0.3]
        \draw[thick, color = red] (0,0) -- (0.75,0);
        
        \filldraw[color = red] (0,0) circle (4pt);
        \filldraw[color = red] (0.75,0) circle (4pt);
    \end{tikzpicture}%
}

\newcommand{\horizontaldimerblue}{%
    \begin{tikzpicture}[baseline=-0.5ex, scale=0.3]
        \draw[thick, color = blue] (0,0) -- (0.75,0);
        
        \filldraw[color = blue] (0,0) circle (4pt);
        \filldraw[color = blue] (0.75,0) circle (4pt);
    \end{tikzpicture}%
}

\newcommand{\singlenndimer}{%
    \begin{tikzpicture}[baseline=0.2ex, scale=0.3]
        \draw[thick, color = black] (0,0) -- (0.75,0.75);

        \filldraw[color = black] (0,0) circle (4pt);
        \filldraw[color = black] (0,0.75) circle (4pt);
        \filldraw[color = black] (0.75,0) circle (4pt);
        \filldraw[color = black] (0.75,0.75) circle (4pt);
    \end{tikzpicture}%
}

\newcommand{\bluesinglenndimer}{%
    \begin{tikzpicture}[baseline=0.2ex, scale=0.3]
        \draw[thick, color = blue] (0,0) -- (0.75,0.75);

        \filldraw[color = blue] (0,0) circle (4pt);
        \filldraw[color = black] (0,0.75) circle (4pt);
        \filldraw[color = black] (0.75,0) circle (4pt);
        \filldraw[color = blue] (0.75,0.75) circle (4pt);
    \end{tikzpicture}%
}

\newcommand{\redsinglenndimer}{%
    \begin{tikzpicture}[baseline=0.2ex, scale=0.3]
        \draw[thick, color = red] (0,0) -- (0.75,0.75);

        \filldraw[color = red] (0,0) circle (4pt);
        \filldraw[color = black] (0,0.75) circle (4pt);
        \filldraw[color = black] (0.75,0) circle (4pt);
        \filldraw[color = red] (0.75,0.75) circle (4pt);
    \end{tikzpicture}%
}

\newcommand{\nnleftdimer}{%
    \begin{tikzpicture}[baseline=0.85ex, scale=0.3]
        \draw[thick, color = black] (0,0) -- (0,0.8);
        \draw[thick, color = black] (0.8,0.8) -- (0,1.6);

        \filldraw[color = black] (0,0) circle (4pt);
        \filldraw[color = black] (0,0.8) circle (4pt);
        \filldraw[color = black] (0.8,0) circle (4pt);
        \filldraw[color = black] (0.8,0.8) circle (4pt);
        \filldraw[color = black] (0,1.6) circle (4pt);
        \filldraw[color = black] (0.8,1.6) circle (4pt);
    \end{tikzpicture}%
}

\newcommand{\nnrightdimer}{%
    \begin{tikzpicture}[baseline=0.85ex, scale=0.3]
        \draw[thick, color = black] (0,0) -- (0.8,0.8);
        \draw[thick, color = black] (0,0.8) -- (0,1.6);

        \filldraw[color = black] (0,0) circle (4pt);
        \filldraw[color = black] (0,0.8) circle (4pt);
        \filldraw[color = black] (0.8,0) circle (4pt);
        \filldraw[color = black] (0.8,0.8) circle (4pt);
        \filldraw[color = black] (0,1.6) circle (4pt);
        \filldraw[color = black] (0.8,1.6) circle (4pt);
    \end{tikzpicture}%
}

\newcommand{\stagdimersecond}{%
    \begin{tikzpicture}[baseline=0.85ex, scale=0.3]
        \draw[thick, color = black] (0.0,0) -- (0.0,0.8);
        \draw[thick, color = black] (0.8,0.8) -- (0.8,1.6);

        \filldraw[color = black] (0,0) circle (4pt);
        \filldraw[color = black] (0,0.8) circle (4pt);
        \filldraw[color = black] (0.8,0) circle (4pt);
        \filldraw[color = black] (0.8,0.8) circle (4pt);
        \filldraw[color = black] (0,1.6) circle (4pt);
        \filldraw[color = black] (0.8,1.6) circle (4pt);
    \end{tikzpicture}%
}

\newcommand{\twonndimersfirst}{%
    \begin{tikzpicture}[baseline=0.85ex, scale=0.3]
        \draw[thick, color = black] (0.0,0) -- (0.8,0.8);
        \draw[thick, color = black] (0.0,0.8) -- (0.8,1.6);

        \filldraw[color = black] (0,0) circle (4pt);
        \filldraw[color = black] (0,0.8) circle (4pt);
        \filldraw[color = black] (0.8,0) circle (4pt);
        \filldraw[color = black] (0.8,0.8) circle (4pt);
        \filldraw[color = black] (0,1.6) circle (4pt);
        \filldraw[color = black] (0.8,1.6) circle (4pt);
    \end{tikzpicture}%
}

\newcommand{\twonndimersfirstblue}{%
    \begin{tikzpicture}[baseline=0.85ex, scale=0.3]
        \draw[thick, color = blue] (0.0,0) -- (0.8,0.8);
        \draw[thick, color = blue] (0.0,0.8) -- (0.8,1.6);

        \filldraw[color = blue] (0,0) circle (4pt);
        \filldraw[color = blue] (0,0.8) circle (4pt);
        \filldraw[color = black] (0.8,0) circle (4pt);
        \filldraw[color = blue] (0.8,0.8) circle (4pt);
        \filldraw[color = black] (0,1.6) circle (4pt);
        \filldraw[color = blue] (0.8,1.6) circle (4pt);
    \end{tikzpicture}%
}

\newcommand{\twonndimersfirstred}{%
    \begin{tikzpicture}[baseline=0.85ex, scale=0.3]
        \draw[thick, color = red] (0.0,0) -- (0.8,0.8);
        \draw[thick, color = red] (0.0,0.8) -- (0.8,1.6);

        \filldraw[color = red] (0,0) circle (4pt);
        \filldraw[color = red] (0,0.8) circle (4pt);
        \filldraw[color = black] (0.8,0) circle (4pt);
        \filldraw[color = red] (0.8,0.8) circle (4pt);
        \filldraw[color = black] (0,1.6) circle (4pt);
        \filldraw[color = red] (0.8,1.6) circle (4pt);
    \end{tikzpicture}%
}

\usepackage{hyperref} 
\hypersetup{
    colorlinks=true,   
    linkcolor=blue,    
    urlcolor=blue,     
    citecolor=blue,    
    allcolors=blue,    
}

\begin{document}

\title{Preparation of a Quantum Spin Liquid in Non-Hermitian Quantum Dimer Models and Rydberg Arrays}

\author{Shashwat Chakraborty}
\affiliation{Department of Physics and Institute for Condensed Matter Theory, University of Illinois Urbana-Champaign, Urbana, IL 61801, USA}
\author{Taylor L. Hughes}
\email{hughest@illinois.edu}
\affiliation{Department of Physics and Institute for Condensed Matter Theory, University of Illinois Urbana-Champaign, Urbana, IL 61801, USA}

\date{\today}

\begin{abstract}

We identify an unconventional form of the non-Hermitian skin effect that occurs not in position space but in many-body Fock space, which we call the Fock space skin effect (FSSE). Using quantum dimer models, we characterize FSSE analytically and numerically, and propose a concrete route toward its realization in Rydberg atom arrays. The dimer constraint is enforced through Rydberg gadgets employing the blockade mechanism, while directional reservoirs generate non-Hermitian flipping amplitudes. We show that FSSE enables the preparation of gapped spin liquid states, and in particular, we demonstrate how a Rydberg geometry realizing a square lattice quantum dimer model with next-nearest neighbor dimers can be driven by non-Hermiticity into an exact spin liquid ground state. Our results establish Fock-space non-Hermiticity as a powerful principle for engineering exotic quantum phases and dynamical state-preparation protocols.

\end{abstract}

\maketitle

Non-unitary dynamics are a central framework for describing open and monitored quantum systems \cite{Fisher2023, Potter2022, Altman2021, Chertkov2023,Ashida2020}. These systems can exhibit rich phenomenology including exotic topological phases \cite{Leykam2017, Kawabata2019} and entanglement phase transitions \cite{Li2018,Skinner2019,Chan2019,Choi2020,Bao2020,Gullans2020,Weinstein2022}. One category of non-unitary dynamics are systems that can be described by non-Hermitian Hamiltonians including quantum systems subject to dissipation \cite{Feshbach1954,Rotter2009} or continuous measurement with postselection \cite{Ueda1990,Dalibard1992,Dum1992,Plenio1998,Daley2014}. Such systems can exhibit phenomena having no Hermitian analog, such as the non-Hermitian skin effect where an extensive number of eigenstates become localized near the boundary of a system \cite{Lee2016, Lin2023, Yao2018, Zhang2022}. 

The phenomenology of the non-Hermitian skin effect generalizes to the domain of intrinsically interacting many-body systems \cite{jacopo1, jacopo2}, where it is characterized by the localization of states not at particular spatial position in the system, but instead on a particular state in the the many-body Fock space \cite{fock_space_NHSE}. We will call such localization in the many-body Fock space the Fock space skin effect (FSSE). 

In parallel, it has been shown that dissipation-driven dynamics can lead to the preparation\cite{Molignini2024} and stabilization\cite{Yang2022} of quantum states of interest. Here we ask the important question: can non-unitary dynamics, and the FSSE in particular, be leveraged to robustly prepare \emph{exotic} quantum states, such as a topological quantum spin liquid? Quantum spin liquids (QSLs) are highly entangled states of matter that elude conventional symmetry-breaking descriptions \cite{Savary2016,Wen2017,Sachdev2018,Balents2010}. The experimental search for QSLs remains a central pursuit in condensed matter physics due to their fundamental interest and potential applications in quantum memory\cite{Dennis2002} and topological quantum computation \cite{Kitaev2003,Kitaev2006,Nayak2008}.
Our study reveals that introducing nonreciprocal couplings into a many-body Hamiltonian can be leveraged to prepare and stabilize a QSL state.
\par 
To this end, here we study the effect of nonreciprocity on quantum dimer models (QDMs).  QDMs offer paradigmatic examples of spin-liquid phenomenology including resonating valence bond (RVB) physics and fractionalized excitations in two-dimensional lattices\cite{RK_QDM_original}. We study the Fock space skin effect in non-Hermitian QDMs and show that by introducing nonreciprocity in the Yao-Kivelson model \cite{Yao_2012}, we can stabilize its gapped spin liquid phase. 

Furthermore, QDMs on square and triangular lattices can be realized on a Rydberg atom array with Rydberg gadgets\cite{rydberg_atom_QDM}, where Rydberg gadgets are auxialliary atoms introduced into the array to impose nontrivial constraints such as the dimer constraint. Here we extend the framework described in Ref. \cite{rydberg_atom_QDM} to include non-Hermiticity by coupling the atoms to directional reservoirs \cite{chiral_res1, chiral_res2}. Our extension provides a methodology for generating the FSSE phenomenon on a Rydberg atom array. Moreover, we utilize FSSE to develop a dynamical scheme for the preparation of a gapped spin-liquid state on a Rydberg atom array.

\begin{figure}[t]
    \centering
    \includegraphics[width=0.95\linewidth]{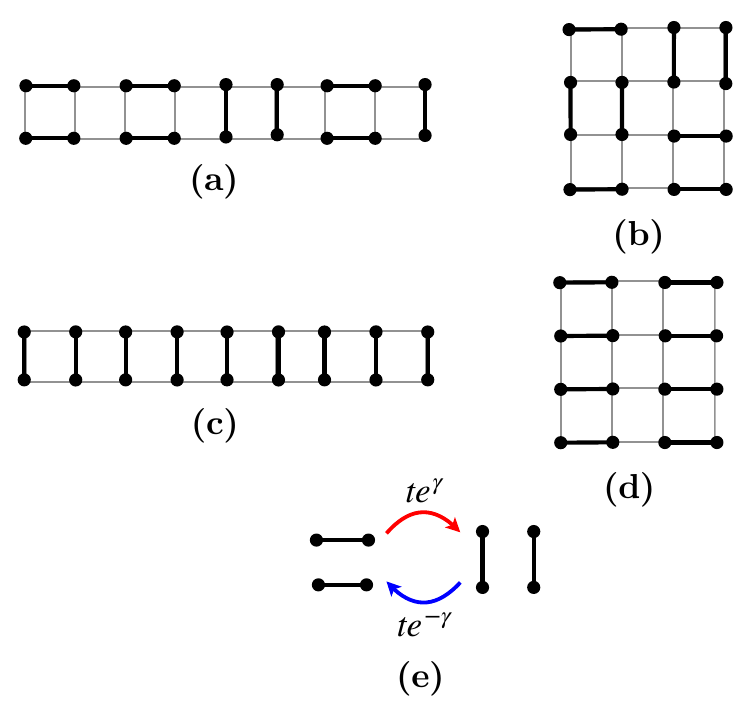}
    \caption{(a) and (b) show examples of close-packed dimer coverings on ladder and square lattice geometries. (c) and (d) show examples of columnar configurations in the ladder and square lattice geometries, respectively. (e) pictorially shows the nonreciprocal flipping amplitudes in the kinetic term of the non-Hermitian QDM (Eq. \ref{eq:T_NH}).} 
    \label{fig:models}
\end{figure}
\textit{QDM on a square lattice.---} The Rokhsar-Kivelson model describes dimers residing on the links of a lattice \cite{RK_QDM_original} such that each site of the lattice is touched by exactly one dimer. Figures \ref{fig:models}{\color{blue}(a-d)} show examples of valid close-packed dimer coverings on square lattice and ladder geometries. The Hamiltonian for the square lattice quantum dimer model involves a plaquette-flip operator, $\hat{T}$, and a dimer-dimer repulsion term, $\hat{V}$, given by 
\begin{equation}
\label{eq:RK_QDM}
\begin{split}
    \hat{H}_{RK} &= \hat{T} + \hat{V}, \\
    \hat{T} &= -t \sum_{\plaquette} \left( \ket{\vplaquette}\bra{\hplaquette} + \ket{\hplaquette}\bra{\vplaquette}\right),\\
    \hat{V} &= V \sum_{\plaquette} \left( \ket{\vplaquette}\bra{\vplaquette} + \ket{\hplaquette}\bra{\hplaquette} \right),
\end{split}
\end{equation}
where $t$ is the flipping amplitude, and $V$ is the repulsion between parallel dimers. The Hilbert space of the RK model consists of different sectors characterized by their topological winding numbers. Following the convention where a dimer is represented by an arrow that goes from sublattice A to B on the square lattice, we can define the winding number $W_x$ ($W_y$) as the number of dimers pointing right (up) minus the number of dimers pointing left (down). The point $t=V$ in the phase diagram of the above QDM is known as the RK point. At the RK point, the ground state of $\hat{H}_{RK}$ in each winding number sector is the equal superposition of all dimer coverings in that sector. 

To study the effects of non-Hermiticity we apply nonreciprocal flipping in the RK model to generate an  example of the FSSE where the ground state localizes on  particular dimer configurations. The plaquette-flip term in the QDM Hamiltonian in Eq. \ref{eq:RK_QDM} with non-reciprocal flipping, as illustrated in Figure \ref{fig:models}{\color{blue}(e)}, changes to 
\begin{equation}
\label{eq:T_NH}
\begin{split}
    \hat{T}_{NH} = -t \sum_{\plaquette} \left( e^{\gamma}\ket{\vplaquette}\bra{\hplaquette} + e^{-\gamma}\ket{\hplaquette}\bra{\vplaquette}\right),
\end{split} 
\end{equation}
which renders the new Hamiltonian as $\hat{H}_{NH} = \hat{T}_{NH} + \hat{V}$. Under a simple similarity transformation, the non-Hermitian Hamiltonian ($\hat{H}_{NH}$) can be mapped back to its Hermitian counterpart (Eq. \ref{eq:RK_QDM}). Implementing the similarity transformation is equivalent to introducing a new bi-orthogonal basis in which $\hat{H}_{NH}$ transforms into $\hat{H}$. The bi-orthogonal basis in which $\hat{H}_{NH}$ becomes Hermitian can be constructed by introducing $\ket{\horizontaldimerred} \equiv e^{\gamma/2}\ket{\horizontaldimer}$ and $\bra{\horizontaldimerblue} \equiv e^{-\gamma/2} \bra{\horizontaldimer}$. In this new bi-orthogonal basis, the kinetic term $\hat{T}_{NH}$ of the Hamiltonian $\hat{H}_{NH}$ transforms into:
\begin{equation}
\begin{split}
    \hat{T}_{NH} = -t \sum_{\plaquette} \left( \ket{\vplaquettered}\bra{\hplaquette} + \ket{\hplaquette}\bra{\vplaquetteblue}\right).
\end{split} 
\end{equation}

The ground state of $\hat{H}_{NH}$ at the RK point $t=V$ can be obtained analytically just as in the Hermitian QDM, but with a key difference. The ground state in a given winding number sector $\mathcal{W}$ is no longer the equal-weight superposition of all dimer coverings, $\mathcal{C}$, in that sector. Instead, each basis state in the superposition now has a weight depending on the number of horizontal dimers\footnote{This is not a unique basis choice. The other alternatives include multiplying the kets and bras with vertical dimers by $e^{\mp \gamma/2}$ or symmetrically scaling the horizontal and vertical dimers such that the overall phase cancels out the nonreciprocity. This would change the weighting structure in the associated ground state.} in the configuration, i.e., 
\begin{equation}
\label{eq:RK_state_NH}
\ket{\psi^{NH}_{\mathcal{W}}} = \sum_{\mathcal{C}\in \mathcal{W}} e^{n_{h}(\mathcal{C})\gamma/2}\ket{\mathcal{C}},
\end{equation}
where $n_{h}(\mathcal{C})$ represents the number of horizontal dimers in the configuration. From the form of the weights entering the ground state $\ket{\psi^{NH}_{\mathcal{W}}}$ we expect it to exhibit the FSSE, i.e., for $\gamma>0$ ($\gamma<0$), and a large system size, $\ket{\psi^{NH}_{\mathcal{W}}}$ is exponentially localized at the configuration with the largest number of horizontal (vertical) dimers. For instance, in the (0,0) winding number sector, the configurations with the largest number of horizontal or vertical dimers are the columnar configurations (see Figures \ref{fig:models}{\color{blue}(c,d)} for examples).

 For a detailed analysis, we consider two geometries for the non-Hermitian RK model: the square lattice and the ladder. The advantage of considering a ladder system in addition to the square lattice is two-fold.  First, the computational complexity of numerically studying the quantum dimer model on a ladder system is much smaller than that of its square-lattice analogue. Second, the ladder geometry is easier to experimentally realize on quantum simulators, such as a Rydberg atom array. 
 
\begin{figure}[t]
    \centering
    \includegraphics[width=0.9\linewidth]{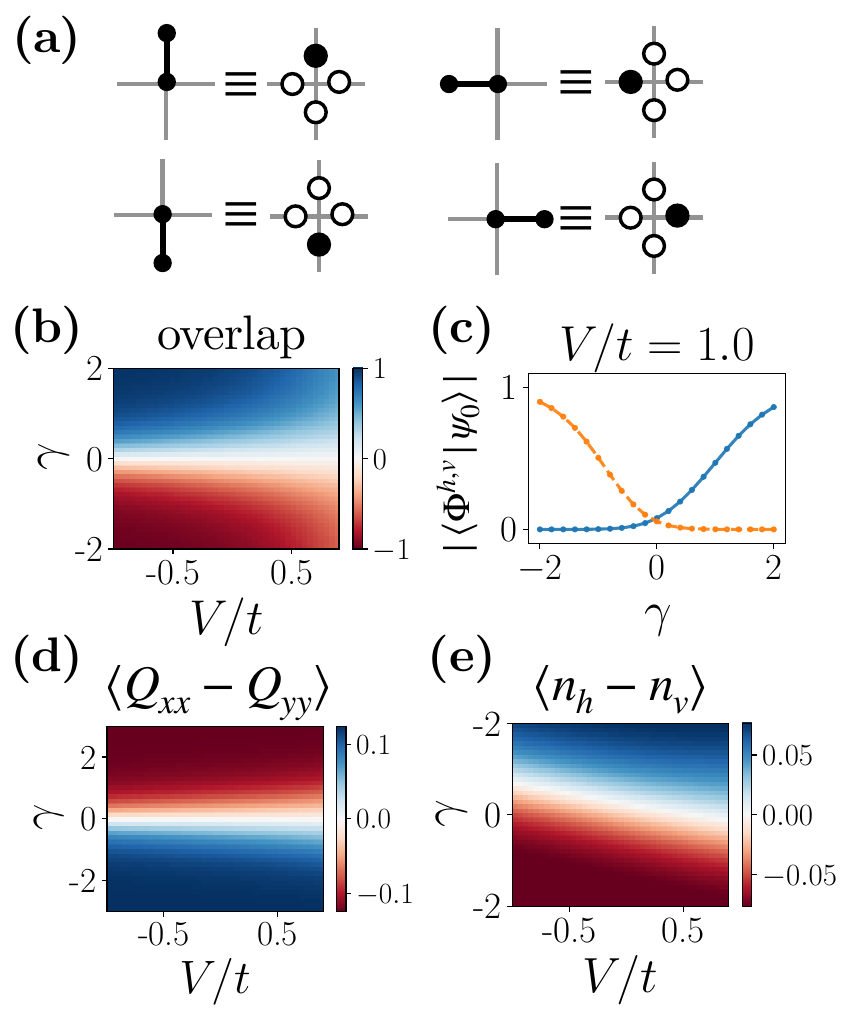}
    \caption{(a) shows the mapping between dimer configurations (left side of identities) to hardcore bosons living on the links of the lattice (right side of identities). {(b) shows the columnar asymmetry, as defined in the main text, of the ground state $\ket{\psi_0(V,\gamma)}$ for the square lattice (system size: $6\times 6$). (c) shows the columnar overlaps of the ground state $\ket{\psi_0(V=t,\gamma)}$ (in the (0,0) winding number sector) for the ladder geometry (system size: $16\times 2$). The blue (orange) curve in (c) corresponds to the overlap of $\ket{\psi_0}$ with the columnar state $\ket{\Phi^h}$ ($\ket{\Phi^v}$) with horizontal (vertical) dimers. (d) shows the average quadrupole moment $\langle Q_{xx} - Q_{yy} \rangle$ in the ground state $\ket{\psi_0(V,\gamma)}$ for the square lattice geometry. (e) shows the asymmetry between horizontal and vertical dimers in the ladder geometry. The asymmetry is non-vanishing even for $\gamma=0$ because the ladder model lacks $C_4$ rotation symmetry. Finite $\gamma$ can tune the asymmetry back to a vanishing point.}}
    \label{fig:numerics}
\end{figure}

 The ground state of the Hermitian QDM on the square lattice lies in the (0,0) winding number sector for $V<t$, and we will restrict ourselves to this sector for all subsequent calculations \footnote{The parameter regime $t>V$ is not of interest because the ground state in this regime lies in the staggered sector, which contains configurations with no flippable plaquettes.}. We denote the ground state of $\hat{H}_{NH}$ for some given $V/t,\gamma$ as $\ket{\psi_{0}(V,\gamma)}$. As we have established above, the FSSE in the (0,0)-sector ground state is expected to manifest as the localization at the columnar configurations. To test this, let us define two reference states $\ket{\Phi^h}$ and $\ket{\Phi^v}$ as equal-weight superpositions of the translation-related columnar configurations with horizontal and vertical dimers, respectively (see Figure \ref{fig:models}{\color{blue}(d)} for an example columnar configuration in the superposition $\ket{\Phi^h}$). We will refer to $\ket{\Phi^h}$ and $\ket{\Phi^v}$ as columnar states. Figure \ref{fig:numerics}{\color{blue}(b)} shows $|\braket{\Phi^{h}|\psi_0(V,\gamma)}|-|\braket{\Phi^v|\psi_0(V,\gamma)}|$, i.e.,  the asymmetry in the overlap of the ground state $\ket{\psi_0(V,\lambda)}$ with the horizontal and vertical columnar states for $V/t\in [-1,1]$ and $\gamma \in (-2,2)$.  A nonzero columnar asymmetry for $\gamma\neq 0$ indicates the onset of FSSE. Note that $\ket{\Phi^h}$ ($\ket{\Phi^v}$) is preferred when $\gamma>0$ ($\gamma<0$), as expected. 

For a ladder system, once again, we restrict ourselves to the $(0,0)$ winding number sector since the ground state lies in this sector for $V<t$. Figure \ref{fig:numerics}{\color{blue}(c)} shows columnar overlaps at $V/t = 1$, and the asymmetry between the two overlaps indicates FSSE. For  the ladder geometry, we also show the horizontal-vertical asymmetry parameter as $\langle{n_{h} - n_{v}}\rangle/L^2$ (see Figure \ref{fig:numerics}{\color{blue}(e)}), as it exhibits an interesting phenomenon. Here, $\langle n_{h}\rangle$ ($\langle n_{v}\rangle$) represents the average number of horizontal (vertical) dimers in the ground state $\ket{\psi_0(V,\lambda)}$ of $\hat{H}_{NH}$. Note that, unlike in the square lattice, the asymmetry is not zero everywhere on the $\gamma = 0$ line. This is because the ladder geometry does not possess a $C_4$ symmetry. Interestingly, by turning on a nonzero $\gamma$, we can make the ground state symmetric in the sense it has an equal number of horizontal and vertical dimers on average. Therefore, a potential application of FSSE could be to endow the ground state of a Hamiltonian with a symmetry that is absent in the Hamiltonian itself. 

In parallel, by using a hardcore bosonic (h.c.b) representation of the RK model with open boundaries\cite{hardcoreboson_QDM}, we can provide an interpretation of this FSSE as the generation of a quadrupole moment in the system\cite{jacopo1}. The Hamiltonian for the RK model can be expressed in terms of hardcore bosonic degrees of freedom living on the links of the square lattice, as follows:
\begin{equation}
\begin{split}
    H^{\text{h.c.b}}_{NH} &= -t\sum_{\boldsymbol{R}} \bigg(e^{-\gamma}\,b^{\dagger}_{\boldsymbol{R}- \hat{e}_1/2}  b^{\dagger}_{\boldsymbol{R}+ \hat{e}_1/2}  b_{\boldsymbol{R}- \hat{e}_2/2} b_{\boldsymbol{R}+ \hat{e}_2/2} \\
    &\quad + e^{\gamma}\, b^{\dagger}_{\boldsymbol{R}-\hat{e}_2/2}  b^{\dagger}_{\boldsymbol{R}+ \hat{e}_2/2}  b_{\boldsymbol{R}- \hat{e}_1/2} b_{\boldsymbol{R}+ \hat{e}_1/2}\bigg)\\
    &\quad + V\sum_{\boldsymbol{R}} \sum_{i=1}^2 b^{\dagger}_{\boldsymbol{R}-\hat{e}_i/2}  b_{\boldsymbol{R}- \hat{e}_i/2} 
    b^{\dagger}_{\boldsymbol{R}+ \hat{e}_i/2}  b_{\boldsymbol{R}+ \hat{e}_i/2},
\end{split}
\end{equation}
where $\boldsymbol{R}$ runs over the sites of the dual lattice, $\hat{e}_i$ ($i = 1,2$) is the unit vector along the $i$-th direction, and $[b_{\boldsymbol{r}},b^\dagger_{\boldsymbol{r}}] = 1-2b^\dagger_{\boldsymbol{r}} b_{\boldsymbol{r}}$. The mapping between dimers and hardcore bosons is illustrated in Figure \ref{fig:numerics}{\color{blue}(a)}. 

The Hamiltonian conserves charge, $N_b=\sum_{\boldsymbol{r}}b^{\dagger}_{\boldsymbol{r}} b_{\boldsymbol{r}}$, as well as the dipole moment, ${\bf{P}}^{(1)} = \sum_{\boldsymbol{r}} \boldsymbol{r} \,b^{\dagger}_{\boldsymbol{r}} b_{\boldsymbol{r}}$, where $\boldsymbol{r}$ runs over the links of the square lattice. The dimer constraint at a site $\boldsymbol{R}$ translates to $\sum_{\boldsymbol{r} \sim \boldsymbol{R}} b^\dagger_{\boldsymbol{r}} b_{\boldsymbol{r}} = 1$, where the sum runs over links connected to $\boldsymbol{R}$. The similarity transformation that removes the nonreciprocity from $H^{\text{h.c.b}}_{NH}$ is given by $S = \exp (\gamma(Q_{xx} - Q_{yy}))$, where
\begin{equation}
    Q_{xx}  = \frac{1}{L^2}\sum_{\boldsymbol{r}} r^2_x \,b^\dagger_{\boldsymbol{r}} b_{\boldsymbol{r}}, \text{ and }Q_{yy}  = \frac{1}{L^2}\sum_{\boldsymbol{r}} r^2_y \,b^\dagger_{\boldsymbol{r}} b_{\boldsymbol{r}},
\end{equation} are components of the bosonic quadrupole moment (assuming open boundary conditions). For an explicit derivation that $S$ removes the nonreciprocity from $H^{\text{h.c.b}}_{NH}$, see the supplementary material \cite{supp}. Figure \ref{fig:numerics}{\color{blue}(d)} shows the behavior of the quadrupole moment $\braket{Q_{xx} - Q_{yy}}$ as we vary $V/t$ and $\gamma$. Evidently, for $\gamma \neq 0$, we observe a nonzero quadrupole moment generated in the system. This behavior is in agreement with the general results of Ref. \cite{jacopo1} for systems obeying non-Hermitian, multipole-conserving dynamics.

\textit{Gapped spin liquid state in square lattice QDM.---} Thus far, we have studied some pedagogical examples of FSSE. We now develop a scheme utilizing FSSE to dynamically prepare a gapped spin liquid state on a square lattice. In Ref. \cite{Yao_2012}, Yao and Kivelson showed that a gapped spin-liquid state can be achieved on a square lattice dimer model by allowing for next-nearest-neighbor (NNN) dimer formation. The model considered in Ref \cite{Yao_2012} includes the following terms involving NNN dimers on top of the RK Hamiltonian (Eq. \ref{eq:RK_QDM}): 
\begin{equation}
\label{eq:YK_QDM}
    \begin{split}
        \hat{T}_1& = -t'\sum_{\left\{\adjplaquettes\right\}}\left(\bigg|\nnleftdimer \bigg\rangle\bigg\langle\nnrightdimer \bigg|+\bigg|\nnrightdimer\bigg\rangle\bigg\langle\nnleftdimer \bigg|\right),\\
        \hat{V}_1 &= V'\sum_{\left\{\adjplaquettes\right\}}\left(\bigg|\nnleftdimer \bigg\rangle\bigg\langle\nnleftdimer \bigg|+\bigg|\nnrightdimer\bigg\rangle\bigg\langle\nnrightdimer \bigg|\right),\\
        \hat{T}_2&=  -t''\sum_{\left\{\adjplaquettes\right\}} 
    \left(\bigg|\twonndimersfirst \bigg\rangle\bigg\langle\stagdimersecond\bigg| + \bigg|\stagdimersecond \bigg\rangle\bigg\langle\twonndimersfirst \bigg|\right),\\
    \hat{V}_2 &=V''\sum_{\left\{\adjplaquettes\right\}} 
\left(\frac{1}{\lambda} \bigg|\twonndimersfirst \bigg\rangle\bigg\langle\twonndimersfirst \bigg| + \lambda\bigg|\stagdimersecond \bigg\rangle\bigg\langle\stagdimersecond \bigg|\right).
    \end{split}
\end{equation}
A sum over all possible symmetry related configurations on pairs of adjacent plaquettes is implied by $\left\{\adjplaquettes\right\}$. From here on, we will refer to the Hamiltonian obtained by adding all the operators in Eq. \ref{eq:YK_QDM} to $\hat{H}_{RK}$ as the Yao-Kivelson Hamiltonian. The generalized RK point of the Yao-Kivelson model appears at $t=V$ (RK point of $\hat{H}_{RK}$), $t'=V'$, and $t'' = V''$. At the generalized RK point, the ground state is:
\begin{equation}
    \ket{\psi_\lambda} = \sum_{\mathcal{C}} \lambda^{n_s(\mathcal{C})/2} \ket{\mathcal{C}},
\end{equation}
where $n_s(\mathcal{C})$ represents the number of NNN dimers in the dimer covering $\mathcal{C}$. The state $\ket{\psi_\lambda}$ can be shown to be a gapped spin-liquid state when the contribution of basis states with NNN dimers, quantified by $\lambda$, is small. The proof involves expressing the partition function in terms of \textit{interacting} Grassmann fields, and mapping it to the massive Thirring model. The massive Thirring model can be solved exactly using the Bethe ansatz to obtain the mass gap as $m(\lambda) \approx m_0 e^{-g/v_F \pi}$ where $g = 2\lambda^2$ and $v_F = 2$ \cite{Yao_2012}. However, the aforementioned mapping to the Thirring model holds only for small $\lambda$. This means that the state $\ket{\psi_\lambda}$ is guaranteed to be a gapped spin liquid state only for small $\lambda$. We now show that the parameter $\lambda$, and hence, the mass gap can be tuned by introducing non-reciprocal flipping amplitudes into the model. 
\par
In the previous examples, we made the kinetic term in $H_{RK}$ non-Hermitian. In contrast, here we introduce nonreciprocal flipping amplitudes only in $\hat{T}_2$ (Eq. \ref{eq:YK_QDM}) and leave the other terms of the Yao-Kivelson Hamiltonian unchanged, i.e., 
\begin{equation}
    \label{eq:T2NH}
    \hat{T}_{2,NH} =  -t''\sum_{\left\{\adjplaquettes\right\}} 
    \left(e^{-\gamma}\bigg|\twonndimersfirst \bigg\rangle\bigg\langle\stagdimersecond\bigg| + e^{\gamma} \bigg|\stagdimersecond \bigg\rangle\bigg\langle\twonndimersfirst \bigg|\right),
\end{equation}
where the parameter $\gamma$ quantifies the non-reciprocity in the Hamiltonian. The non-reciprocity can be removed from the Hamiltonian by performing a similarity transformation as before. The bi-orthogonal basis that renders $\hat{T}_{2,NH}$ (Eq. \ref{eq:T2NH}) Hermitian is given by $\ket{\redsinglenndimer} = e^{-\gamma/2} \ket{\singlenndimer}$, $\bra{\bluesinglenndimer} = e^{\gamma/2} \bra{\singlenndimer}$. The flipping term $\hat{T}_{2,NH}$ upon making the similarity transformation, takes the form:
\begin{equation}
    \label{eq:T2NH_new}
    \hat{T}_{2,NH} =  -t''\sum_{\left\{\adjplaquettes\right\}} 
    \left(\bigg|\twonndimersfirstred \bigg\rangle\bigg\langle\stagdimersecond\bigg| + \bigg|\stagdimersecond \bigg\rangle\bigg\langle\twonndimersfirstblue \bigg|\right).
\end{equation}
For $\gamma>0$, the new ground state $\ket{\psi_{\text{new}}}$ has a smaller contribution of basis states with NNN dimers, i.e.,
\begin{equation}
\label{eq:new_GS_YK}
    \ket{\psi_{\text{new}}} = \sum_{\mathcal{C}} (\lambda e^{-\gamma})^{n_s(\mathcal{C})/2} \ket{\mathcal{C}} = \ket{\psi_{\lambda'}}, 
\end{equation}
where $\lambda' = \lambda e^{-\gamma}$. By making the value of $\gamma$ large, we can make the contribution of basis states with NNN dimers arbitrarily small. Therefore, starting from a state $\ket{\psi_{\lambda}}$ with large $\lambda$, and then introducing non-Hermitian couplings,  we can approach a spin-liquid state as we increase the parameter $\gamma$.    
\par
\begin{figure}[t]
    \centering
    \includegraphics[width=\linewidth]{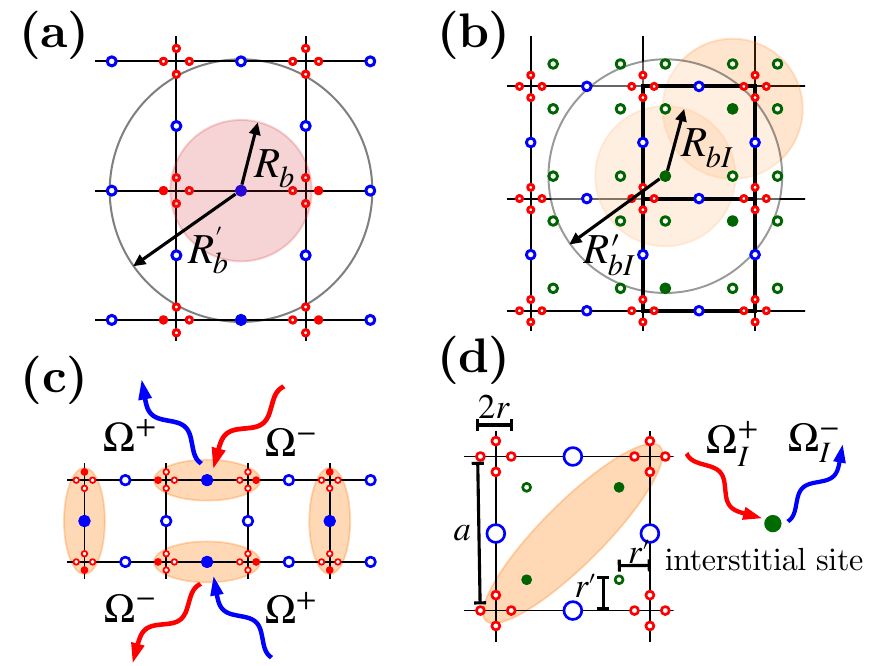}
    \caption{Rydberg atom geometries for quantum dimer models. (a) shows the geometry optimal for the realization of the QDM on a square lattice. The link atoms (blue) are of the same species as the gadget atoms (red). The lattice spacing is taken such that each link atom blockades six gadget atoms, as shown. Here $R'_b$ is the radius of the extended interaction range. (b) shows the Rydberg-atom configuration that achieves the Yao-Kivelson model. This geometry includes interstitial atoms (green) that encode NNN dimers. Here $R_{bI}$ is the blockade radius and $R'_{bI}$ is the radius of the extended interaction range of the interstitial atoms. (c) shows a valid dimer covering in the ladder geometry. Each dimer comprises an active link atom and two active gadget atoms. The same holds for the square lattice geometry as well. We couple the link atoms only on the horizontal links (or equivalently, vertical links) to directional reservoirs to achieve non-Hermitian couplings in the plaquette-flip term of the effective Hamiltonian. (d) illustrates that two active interstitial atoms on a diagonal encode an NNN dimer on that diagonal. We couple the interstitial atoms to directional reservoirs to achieve the non-Hermitian flipping term in Eq. \ref{eq:T2NH}. The  blue and red wavy  arrows represent the new amplitudes of the $\ket{\text{Gnd}}\bra{\text{Ryd}}$ and $\ket{\text{Ryd}}\bra{\text{Gnd}}$ of the interstitial atoms, respectively. }
    \label{fig:geometries}
\end{figure}

 
\textit{Possible Realization using Rydberg Gadgets.---} In recent years, quantum simulators, such as neutral atom arrays, have emerged as promising platforms for the experimental realization of various topological phases of matter \cite{Semeghini2021}. These platforms offer high programmability that can be utilized in the domains of quantum simulation \cite{Bernien2017,Labuhn2016,Keesling2019,Ebadi2021,Bluvstein2021}, quantum optimization \cite{Ebadi2022}, and quantum computing \cite{Bluvstein2022, Bluvstein2024,Levine2019}. A recent proposal provided a framework to simulate quantum dimer models on square and triangular lattices on a Rydberg atom array \cite{rydberg_atom_QDM}. We start with the arrangements illustrated in Figure \ref{fig:geometries} \cite{rydberg_atom_QDM}, and then develop a protocol for a possible realization of the non-Hermitian QDMs discussed in this work using Rydberg atoms coupled to directional reservoirs \cite{chiral_res1, chiral_res2}. 

The general Hamiltonian describing a Rydberg-atom array is the following:
\begin{equation}
\label{eq:Rb_hamiltonian}
 \hat{H}_{Ry} = \Omega \sum_{n} \hat{\sigma}^x_n - \sum_{n} \Delta_n \,\hat{n}_n + \sum_{n,m}  U_{n,m} \,\hat{n}_n\, \hat{n}_{m},
\end{equation}
where $n,m$ label the atoms in the array, $\Omega$ is the Rabi frequency, and $\Delta_n$ is the detuning of atom $n$. The two possible states of a Rydberg atom are $\ket{\text{Gnd}}$, the ground (inactive) state, and $\ket{\text{Ryd}}$, the Rydberg (active) state. In Eq. \ref{eq:Rb_hamiltonian}, $\hat{\sigma}^x = \ket{\text{Gnd}}\bra{\text{Ryd}}+\ket{\text{Ryd}}\bra{\text{Gnd}}$, and $\hat{n} = \ket{\text{Ryd}}\bra{\text{Ryd}}$. The repulsion between two atoms is short-ranged and van der Waals in nature, i.e., $U_{n,m} \sim 1/|\boldsymbol{x}_n - \boldsymbol{x}_{m}|^6$. Because at large distances, $U_{n,m}$ decays rapidly, it is common practice to make a blockade approximation to truncate the interaction range of $U_{n,m}$ \cite{Jaksch2000FastGates,Lukin2001DipoleBlockade,Gaetan2009RydbergBlockade}. Under the blockade approximation, we replace the potential $U_{n,m}$ by the following finite-ranged potential:
\begin{equation}
    {U}^b_{n,m} = \begin{cases}
        \infty &\text{if }|\boldsymbol{x}_n - \boldsymbol{x}_{m}| \leq R_b\\
        0 &\text{otherwise}
    \end{cases}, 
\end{equation}
\noindent where $R_b$ is called the blockade radius. 

We work in the $\Omega/\Delta <<1$ regime, which enables us to treat the Rabi term perturbatively \cite{rydberg_atom_QDM}. By adjusting the geometry of the system, we can realize a wide variety of models as effective leading-order Hamiltonians in the perturbative expansion of $\hat{H}_{Ry}$ in orders of $\Omega/\Delta$\cite{supp}. We place a Rydberg atom on each link of the square lattice (blue circles in Figure \ref{fig:geometries}{\color{blue}(a)}) and four Rydberg gadgets (red circles in Figure \ref{fig:geometries}{\color{blue}(a)}) near each vertex, which are auxilliary Rydberg atoms that enforce the dimer constraint. Figure \ref{fig:geometries}{\color{blue}a} shows the geometry that realizes the square lattice QDM, in which the presence/absence of a dimer is encoded by the state of an active atom on a lattice link together with two active Rydberg gadgets located near the two vertices connected by that link. In the absence of the Rabi term, the ground state subspace of the Rydberg Hamiltonian consists of all close-packed dimer coverings, provided that the detuning of the gadget atoms is the same as that of the link atoms \cite{supp}. A perturbative expansion of $\hat{H}_{Ry}$ generates the plaquette-flip term, $\hat{T}$, in Eq. \ref{eq:RK_QDM} with a flipping amplitude $\sim \Omega^{12}/\Delta^{11}$ \cite{rydberg_atom_QDM}. 

In contrast,  the nontrivial dimer-dimer repulsion term, $\hat{V}$, in Eq. \ref{eq:RK_QDM} does not appear in the expansion. This can be rectified by adding a van der Waals tail to ${U}^{b}_{nm}$ \cite{supp} (see Figure \ref{fig:geometries}{\color{blue}(a,b)}). We will denote $\tilde{U}_{nm}$ as the potential $U^b_{nm}$ with the tail. A heuristic explanation follows. Without the tail in $\tilde{U}_{nm}$, the dimers on parallel links of a plaquette do not interact. Thus, the energy cost of placing two parallel dimers on a plaquette is zero. In the presence of a tail in $\tilde{U}_{nm}$, atoms on opposite links have a non-trivial repulsion energy. We leverage this non-trivial repulsion between atoms on opposite links to obtain the dimer-dimer repulsion term in the RK Hamiltonian (Eq. \ref{eq:RK_QDM}). 
\par 
The plaquette-flip term obtained in the perturbative expansion of $\hat{H}_{Ry}$ can be made nonreciprocal by coupling the link atoms on horizontal links (or equivalently, vertical links) with directional (chiral) reservoirs. A directional reservoir can be realized as a chiral one-dimensional waveguide, where the coupling to right-moving photons is strong, while that to left-moving photons is suppressed, or vice-versa. For an atom at position $n$ coupled to a directional reservoir, the Rabi term in $\hat{H}_{Ry}$ (Eq. \ref{eq:Rb_hamiltonian}), $\Omega \sigma^x_n$, changes to $\Omega^+ \sigma^+_n + \Omega^{-} \sigma^{-}_{n}$, where $\Omega^{+} \neq \Omega^{-}$ as represented by blue/red wavy arrows in Figures in \ref{fig:geometries}{\color{blue}(c,d)}. We call the Rydberg array Hamiltonian with nonreciprocal plaquette-flip terms $\hat{H}^{NH}_{Ry}$. The new flipping amplitudes in the effective Hamiltonian obtained by perturbatively expanding $\hat{H}^{NH}_{Ry}$ are $(\Omega^{+})^{2}\Omega^{10}/\Delta^{11}$ and $(\Omega^{-})^{2}\Omega^{10}/\Delta^{11}$, which we identify with the flipping amplitudes $te^{\gamma}$ and $te^{-\gamma}$ in Eq. \ref{eq:T_NH}. Therefore, the leading order effective Hamiltonian emerging from $H^{NH}_{Ry}$ maps exactly to the non-Hermitian RK model described in Eq. \ref{eq:T_NH}. We expect the above setup to exhibit FSSE, i.e., localization of the ground state near the columnar configurations with horizontal ($\gamma>0$) or vertical ($\gamma<0$) dimers. We defer an extensive numerical study of the Rydberg atom array to future work.

To generate a spin liquid the geometry must be generalized to include NNN dimers by introducing interstitial atoms (green circles) to encode NNN dimers, as illustrated in Figure \ref{fig:geometries}{\color{blue}b}. We assume that the detuning and Rabi frequency for the interstitial atoms are $\Delta_{I}$ and $\Omega_{I}$, respectively. The low-energy subspace of the diagonal part of $\hat{H}_{Ry}$, i.e., everything except the Rabi term in Eq. \ref{eq:Rb_hamiltonian}, contains the allowed dimer coverings of the Yao-Kivelson model when $\Delta_{I} = 3\Delta/2$. This is because we need the configurations connected by $\hat{T}_{2,NH}$ (Eq. \ref{eq:T2NH}) to all lie in the low energy subspace of the diagonal part of $\hat{H}_{Ry}$. Therefore, the energy of replacing a staggered configuration over two adjacent plaquettes by two NNN dimers on those plaquettes must be equal.

To find the effective low-energy model we again treat the Rabi term perturbatively. The plaquette-flip term, $\hat{T}$, of the RK Hamiltonian involves 12 flips (4 link atoms and 8 gadget atoms). The flip term $\hat{T}_1$ on adjacent plaquettes involves 10 flips (2 interstitial atoms, 2 link atoms, 4 gadget atoms). Therefore, $\hat{T}_1$ appears at tenth order in the perturbative expansion of $\hat{H}_{Ry}$, i.e., $t' \sim \Omega^8 \Omega_{I}^2/(\Delta^{9})$. Similarly, $\hat{T}_2$ appears at the tenth order to render $t'' \sim \Omega^6 \Omega^4_{I}/\Delta^{9}$, since it involves flipping 4 interstitial atoms, 2 link atoms, and 4 gadget atoms. As above, the nontrivial diagonal terms appear as a result of the tail in $\tilde{U}_{nm}$. 

 In order to realize the non-Hermitian Yao-Kivelson model, we couple the interstitial atoms to directional reservoirs. In the presence of directional reservoirs, the Rabi term for an interstitial atom $n$ becomes $\Omega^+_{I} \sigma^+_n + \Omega_{I}^{-} \sigma^{-}_{n}$. We parametrize $\Omega^{\pm}_I$ as $\Omega_Ie^{\mp \gamma}$. The generalized RK point in this setup can be achieved by tuning $\Omega, \Omega_I$, and the interstitial location encoded by $r'$ \cite{supp}. At the generalized RK point, the ground state is given by the analog of Eq. \ref{eq:new_GS_YK} which was written in terms of dimer coverings. By increasing the nonreciprocity parameter $\gamma$, we expect to approach a gapped spin liquid state on the Rydberg atom array. Thus, the Rydberg array geometry we prescribe can be used to experimentally study the phase diagram of the Yao-Kivelson model.

\textit{Conclusions.---}In this article, we uncover an unconventional form of the non-Hermitian skin effect that emerges not in position space but in many-body Fock space, which we term the Fock space skin effect. Focusing on quantum dimer models, we characterize FSSE both analytically and numerically, and outline a route toward its experimental realization in Rydberg atom arrays. The dimer constraint can be enforced via Rydberg gadgets utilizing the blockade mechanism, while non-Hermitian flipping amplitudes arise from selective coupling to directional reservoirs. We show that the Fock space skin effect in the non-Hermitian RK model manifests itself as the generation of a quadrupole moment, when viewed as a dipole-conserving hardcore bosonic Hamiltonian. Our discussion of the realization of non-Hermitian QDMs on a Rydberg atom array provides a route to experimentally observe non-Hermitian skin effect in two-dimensional many-body dipole conserving dynamics.  

Remarkably, FSSE provides a pathway for preparing a gapped spin liquid state on the square lattice. In particular, we showed Rydberg arrays can realize the Yao–Kivelson model, which can be driven by non-Hermiticity into a regime where the ground state is an exact spin liquid. We defer a quantum Monte-Carlo study of the Yao-Kivelson model and its associated Rydberg atom geometry to future work. Beyond these results, our work suggests that FSSE may offer a versatile framework for engineering exotic quantum phases, with potential extensions to other constrained models and to dynamical protocols for state preparation. 

\begin{acknowledgments}
T.L.H. and S.C. acknowledge support from the US Office of Naval Research MURI grant N00014-20-1-2325. 
\end{acknowledgments}

\bibliography{references} 
\end{document}


\title{Supplemental Material for ``Preparation of a Quantum Spin Liquid in Non-Hermitian Quantum Dimer Models and Rydberg Arrays''} 

\author{Shashwat Chakraborty}
\affiliation{Department of Physics and Institute for Condensed Matter Theory, University of Illinois Urbana-Champaign, Urbana, IL 61801, USA}
\author{Taylor L. Hughes}
\email{hughest@illinois.edu}
\affiliation{Department of Physics and Institute for Condensed Matter Theory, University of Illinois Urbana-Champaign, Urbana, IL 61801, USA}

\date{\today}

\maketitle

\section{S1: Atom Arrangements and Effective Hamiltonian for QDMs} \label{sec:derivation}
Recently, it was shown that QDMs on square and triangular lattices can be simulated on Rydberg atom arrays by employing Rydberg gadgets to encode the dimer constraints at each site \cite{rydberg_atom_QDM}. In a Rydberg atom array, neutral atoms are approximated as two-level systems — the ground state $\ket{\text{Gnd}}$ and the highly excited state $\ket{\text{Ryd}}$. This platform offers excellent freedom to configure the geometry of the atoms and to simulate a wide variety of Hamiltonians. Each atom is driven by a laser with Rabi frequency $\Omega$ and detuning $\Delta$. The Hamiltonian for the system is given by:
\begin{equation}
\label{eq:rydberg_hamiltonian}
    \hat{H}_{Ry} = \Omega \sum_{i} \hat{\sigma}^x_i - \sum_{i} \Delta_i \,\hat{n}_i + \sum_{i,j}  U_{i,j} \,\hat{n}_i\, \hat{n}_j
\end{equation} 
where $\sigma^x \equiv \ket{\text{Gnd}}\bra{\text{Ryd}} + \ket{\text{Ryd}} \bra{\text{Gnd}}$, $\hat{n} \equiv \ket{\text{Ryd}} \bra{\text{Ryd}}$, and the index $i$ labels the position $\boldsymbol{x}_i$ of each atom. The interaction $U_{ij}$ is of the van der Waals form:
\begin{equation}
    U_{ij} \sim \frac{1}{|\boldsymbol{x}_i - \boldsymbol{x_j}|^6}
\end{equation}
It is evident from the above equation that the interaction is force repulsive at short distances and its strength decays rapidly at longer distances. This behavior enables us to approximate the interaction neighborhood as a disc of radius $R_b$, known as the \textit{blockade} radius. All interactions within the blockade radius are taken to be infinite. Put differently, the blockade approximation translates into the following condition: $\hat{n}_i \hat{n}_j = 0$ if $|\boldsymbol{x}_i - \boldsymbol{x}_j| < R_b$. We will first review the Rydberg atom geometry that realizes the square lattice QDM \cite{rydberg_atom_QDM}, and then extend this framework to the model proposed by Yao and Kivelson \cite{Yao_2012}.

\subsubsection{Rokhsar-Kivelson Square Lattice QDM}
\begin{figure}
    \centering
    \includegraphics[width=\linewidth]{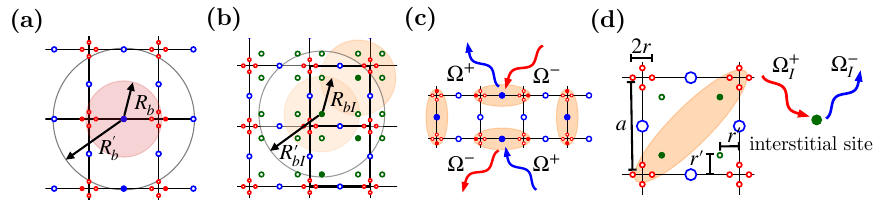}
    \caption{Rydberg atom geometries for quantum dimer models. (a) shows the geometry optimal for the realization of the QDM on a square lattice. The link atoms (blue) are of the same species as the gadget atoms (red). The lattice spacing is taken such that each link atom blockades six gadget atoms, as shown. Here $R'_b$ is the radius of the extended interaction range. (b) shows the Rydberg-atom configuration that achieves the Yao-Kivelson model. This geometry includes interstitial atoms (green) that encode NNN dimers. Here $R'_{b,2}$ is the radius of the extended interaction range for the interstitial atoms. (c) shows a valid dimer covering in the ladder geometry. Each dimer comprises an active link atom and two active gadget atoms. The same holds for the square lattice geometry as well. We couple the link atoms only on the horizontal links (or equivalently, vertical links) to directional reservoirs to achieve non-Hermitian couplings in the plaquette-flip term of the effective Hamiltonian. (d) illustrates that two active interstitial atoms on a diagonal encode an NNN dimer on that diagonal. We couple the interstitial atoms to directional reservoirs to achieve the non-Hermitian flipping term of the non-Hermitian Yao-Kivelson model ($T^{NH}_{2}$ in the main text). The  blue and red wavy  arrows represent the new amplitudes of the $\ket{\text{Gnd}}\bra{\text{Ryd}}$ and $\ket{\text{Ryd}}\bra{\text{Gnd}}$ of the interstitial atoms, respectively.}
    \label{fig:rydberg_geometries}
\end{figure}

To realize the square-lattice QDM, we consider the following geometry: 
\begin{enumerate}
    \item Rydberg atoms are placed on the links of a square lattice (see Figure \ref{fig:rydberg_geometries}a). We will refer to these as \textit{link} atoms. 
    \item The dimer constraint is enforced by placing four Rydberg gadgets near each vertex of the square lattice (see Figure \ref{fig:rydberg_geometries}a). We will refer to these as \textit{gadget} atoms.
    \item We take the lattice spacing such that the blockade region around every link atom includes exactly six gadget atoms nearest to it (see Figure \ref{fig:rydberg_geometries}b). 
\end{enumerate}
Active (inactive) atoms are represented by closed (open) circles. The off-diagonal term of the Hamiltonian,  $\hat{V}=\Omega\sum_{i} \sigma_i^x$, is responsible for the creation or destruction of a dimer. The effective Hamiltonian at leading order $n$ is given by \cite{perturbative_expansion}:
\begin{equation}
    \hat{H}_{\text{eff}} \sim \hat{P}\hat{V} \left(\hat{Q}\frac{1}{E_0 - \hat{Q}\hat{H}_0 \hat{Q}}\hat{Q} \hat{V}\right)^{n-1} \hat{P}
\end{equation}
where $\hat{P}$ is the projector onto the ground subspace of $\hat{H}_0$, and $\hat{Q} = \mathbb{I}- \hat{P}$. Upon perturbatively expanding the Hamiltonian, $\hat{H}_{Ry}$ (Eq. \ref{eq:rydberg_hamiltonian}), we obtain the off-diagonal term of the RK Hamiltonian at leading order. The flipping amplitude $t$ can be shown to be of order $\Omega^{12}/\Delta^{11}$. Unfortunately, the nontrivial diagonal term in the RK model that represents repulsion between parallel dimers does not appear in the perturbative expansion of $\hat{H}_{Ry}$. However, an interesting feature of the gadget model, described above, is that its ground state $\ket{\psi_0}$ approaches the RVB state, $\ket{\text{RVB}} = \sum_{\mathcal{C}} \ket{\mathcal{C}}$, as $\Delta/\Omega$ increases. This can be seen numerically by computing the overlap $\braket{\text{RVB}|\psi_0}$. A rigorous explanation for why the Hamiltonian approaches the RVB state is still missing, since the effective Hamiltonian obtained from the perturbative expansion of $\hat{H}_{Ry}$ does not contain the RK point ($V=t$) of the QDM Hamiltonian.    
\par 
\vspace{2mm}
We now consider a slightly modified version of the blockade approximation. In the previous discussion, we imposed sharp boundaries on the blockade region. In the following, we relax this condition to include small repulsive interactions beyond the blockade region. To demonstrate how the nontrivial dimer-dimer repulsion term in the RK model can arise in the perturbative expansion of $\hat{H}_{Ry}$, we can take, for example, the following definition of $\tilde{U}_{ij}$:
\begin{equation}
    \tilde{U}_{ij} = \begin{cases}
    \infty &\text{for } |\boldsymbol{x}_i - \boldsymbol{x}_j| \leq R_b\\
    u & \text{for }  R_b<|\boldsymbol{x}_i - \boldsymbol{x}_j| \leq R_b'\\
    \end{cases},
\end{equation}
where $R_b' \gtrsim a$. We assume that $u$ is much smaller than $\Delta$. We reiterate that this approximation is made just for the purposes of a simple demonstration. The lack of distance dependence in the potential $U(r)$ beyond $r = R_b$ is unphysical. We include the distance dependence in the potential in the following subsection. The fact that $u$ is much smaller than $\Delta$ implies that the largest energy scale in the system is still $\Delta$. Close-packed dimer configurations no longer form the ground state subspace of the unperturbed Hamiltonian, i.e., the diagonal part of $\hat{H}_{Ry}$ (treating the Rabi term as a perturbation). To see this, let us take two examples. First, when there are two parallel dimers on a plaquette. The repulsion energy in this case is $7u$. The second example is where we have a staggered configuration on two adjacent plaquettes. In this case, the total repulsion energy is $4u$. Clearly, there is a non-trivial energy cost of $3u$ for having parallel dimers on a plaquette, just as in the RK model. Even though the close-packed configurations are not exactly degenerate, they are separated from the rest of the spectrum by a gap of order $\Delta$. This enables us to treat the Rabi term in $\hat{H}_{Ry}$ under the (approximate) degenerate perturbation theory as before. Therefore, from here on, we do not impose a hard cutoff on the extended interaction region, i.e., the extent of $\tilde{U}_{ij}$. 
\par 
\vspace{2mm}
The effective Hamiltonian that one gets from the perturbative expansion of $\hat{H}_{Ry}$ in this case is given by:
\begin{equation}
    H_{\text{eff}} \sim -t \sum_{\plaquette} \left(\ket{\vplaquette}\bra{\hplaquette} + \ket{\hplaquette}\bra{\vplaquette}\right)+ V\sum_{\plaquette} \left(\ket{\vplaquette}\bra{\vplaquette} + \ket{\hplaquette}\bra{\hplaquette}\right),
\end{equation}
where $t \sim \Omega^{12}/\Delta^{11}$ and $V \sim 3u$. The Rabi frequency $\Omega$ can be tuned to achieve the RK point $t=V$. As mentioned above, the interaction range can be extended in a more physical way by including the distance dependence of the interaction potential. In this picture, the radius $R'_b$ can be interpreted as the distance beyond which the interaction strength becomes negligibly small compared to its strength just outside the blockade region.   

\subsubsection{Square Lattice QDM with NNN Dimers}
We note that the detuning of the interstitial atoms $\Delta_{I}$ must be $3\Delta/2$ so that there is no energy cost to replace two staggered NN dimers on adjacent plaquettes by a pair of parallel NNN dimers. The plaquette-flip term in the RK Hamiltonian involves 12 flips. This is because we need to flip the state of four link atoms (blue) and eight gadget atoms (red) to reach an allowed configuration. Similarly, we can calculate the number of flips required for all other terms in the Yao-Kivelson Hamiltonian. 
\begin{equation}
    \begin{split}
        &\vplaquette \leftrightarrow \hplaquette \quad \text{12 flips} \implies t \sim \frac{\Omega^{12}}{\Delta^{11}}\\
        &\nnleftdimer \leftrightarrow \nnrightdimer \quad \text{10 flips}\implies t' \sim \frac{\Omega^{10}}{\Delta^9}\\
        &\twonndimersfirst \leftrightarrow \stagdimersecond \quad \text{10 flips} \implies t'' \sim \frac{\Omega^{10}}{\Delta^9}
    \end{split}
\end{equation}
However, at this point, we are faced with a major roadblock. Getting the right diagonal term to realize the RK point is not possible. The non-trivial diagonal terms that act only on flippable plaquettes do not appear in the perturbative expansion, as mentioned in the previous subsection. 

\vspace{2mm}
Similarly to the previous subsection, we now redo the above calculation using the modified blockade approximation. Before we move on to find the effective Hamiltonian with the modified blockade approximation, we need to make some key assumptions. First, we assume that the three types of atom in the model---gadget, link, and interstitial---are driven by different Rabi frequecies, $\Omega_g,\Omega_e (=\Omega_g),$ and $\Omega_I$, respectively. Second, we assume that interstitial atoms have a larger gap between their ground and excited states. This would enable us to selectively couple the interstitial atoms to the environment without affecting the gadget and link atoms. With these assumptions in mind, we now derive the effective Hamiltonian for the NNN model. 
\par 
\vspace{2mm}
The total repulsion energy for having parallel NNN dimers on a pair of adjacent plaquettes $\left(\twonndimersfirst\right)$ is $5u$. The repulsion energy between a NN dimer and a NNN dimer on two adjacent plaquettes $\left(\nnleftdimer\right)$ is also $5u$. The flipping amplitudes $t$, $t'$, and $t''$ can once again be estimated as follows:
\begin{equation}
    \begin{split}
        &\vplaquette \leftrightarrow \hplaquette \quad \text{12 flips} \implies t \sim \frac{\Omega_e^{4}\Omega_g^8}{\Delta^{11}}\\
        &\nnleftdimer \leftrightarrow \nnrightdimer \quad \text{10 flips}\implies t' \sim \frac{\Omega_e^{2} \Omega_g^{4} \Omega_I^{4}}{\Delta^9}\\
        &\twonndimersfirst \leftrightarrow \stagdimersecond \quad \text{10 flips} \implies t'' \sim \frac{\Omega_e^{2} \Omega_g^{4} \Omega_I^{4}}{\Delta^9}
    \end{split}
\end{equation}
The generalized RK point can be obtained by first tuning $\Omega_e$ and $\Omega_g$ so that $t = 3u$. Then we adjust $\Omega_I$ to satisfy $t' (=t'') = u$. This enables us to realize the generalized RK point of the Yao-Kivelson Hamiltonian with $\lambda = 1$. Note that the ground state of the Yao-Kivelson Hamiltonian at $\lambda = 1$ is not guaranteed to be a spin liquid state. Therefore, we need a way to tune $\lambda$. Adding non-Hermiticity can do the trick. We couple the interstitial atoms to a directional reservoir (also known as a chiral reservoir). This changes the excitation and de-excitation amplitudes to $\Omega^+_I$ and $\Omega^{-}_I \neq \Omega^{+}_I$, respectively. The flipping amplitude $t'$ in the effective Hamiltonian get modified to $\sim (\Omega^{+}_I\Omega^{-}_I)^2\Omega^6/\Delta^9$. The originally hermitian flipping between $\ket{\twonndimersfirst}$ and $\ket{\stagdimersecond}$ becomes non-Hermitian with new flipping amplitudes $t''_{NNN\to \text{stag.}}$ and $t''_{\text{stag.}\to NNN}$ given by $(\Omega^{-}_I)^4\Omega^6/\Delta^9$ and $(\Omega^{+}_I)^4\Omega^6/\Delta^9$, respectively. 
\par
\vspace{2mm}
Now we discuss a more physical approximation. We take the interactions outside the blockade region as distance-dependent. We only need to consider the following interactions. 
\begin{enumerate}
    \item  The distance between a gadget atom and an link atom in a dimer is $a/2+r$. The repulsion at this separation is given by $C_{ge}/(a+r/2)^6$. 
    \item The distance between atoms on opposite links in a parallel dimer configuration is $a$. The repulsion for this configuration is given by $C_{gg}/a^6$ and $C_{ee}/a^6$. 
    \item The next configuration we consider is the staggered configuration of dimers on a pair of adjacent plaquettes. The only non-trivial repulsion is the one between the two gadget atoms closest to each other. The separation between the mentioned gadget atoms is $2\sqrt{r^2+(a/2)^2}$. The repulsion strength is thus given by $C_{gg}/[4(r^2+(a/2)^2)]^3$. 
    \item Next we consider the configuration with a NNN dimer and a NN dimer on a pair of adjacent plaquettes. The only nontrivial interactions are the ones between the interstitial atoms and the active gadget atom closest to them. The relevant distance here is $\sqrt{(a-r')^2+ (r'-r)^2}$. The repulsion strength is $C_{ig}/((a-r')^2+ (r'-r)^2)^3$.
    \item Finally, we compute the repulsion between two parallel NNN dimers on two adjacent plaquettes. The distances involved are $a$ and $\sqrt{(a-2r')^2+4r'^2}$. Their respective repulsion strengths are $C_{ii}/a^6$ and $C_{ii}/((a-2r')^2+4r'^2)^3$. 
\end{enumerate}
The formation energy of an NN dimer is $2C_{ge}/(a+r/2)^6$, and that of an NNN dimer is $C_{ii}/[8(a-2r')^6]$. Let us now explicitly write out the total potential energy of each configuration of interest. In the following, we assume that $C_{gg} = C_{ee} = C_{ge} (=C)$ because the gadget and link atoms are of the same species. We start with the parallel NN dimer configuration on a plaquette the total interaction potential energy of which is given by: 
\begin{equation}
    U(\vplaquette) = U(\hplaquette) = \frac{3C}{a^6}. 
\end{equation}
Note that we subtracted the formation energy of two dimers separate out the dimer-dimer interaction energy. Similarly, the interaction potential energies of different configurations on a pair of adjacent plaquettes are given by:
\begin{align}
U\left(\stagdimersecond\right) &= \frac{C}{[4(r^2+(a/2)^2)]^3},\\
U\left(\nnleftdimer\right) = U\left(\nnrightdimer\right)&= \frac{2C_{ig}}{[(a-r')^2+ (r'-r)^2]^3} + \frac{C_{ii}}{8(a-2r')^6} - \frac{2C}{(a+r/2)^6}\\
U\left(\twonndimersfirst\right) &= \frac{2C_{ii}}{a^6} + \frac{2C_{ii}}{8(a-2r')^6} + \frac{C_{ii}}{[(a-r')^2+ (r'-r)^2]^3}- \frac{4C}{(a+r/2)^6}
\end{align}

We find $r'$ such that the following holds:
\begin{equation}
\label{eq:condition_for_generalized_RK}
U\left(\nnleftdimer\right)=\sqrt{U\left(\twonndimersfirst\right)U\left(\stagdimersecond\right)}
\end{equation}
Eq. \ref{eq:condition_for_generalized_RK} ensures that the RK point in the Yao-Kivelson Hamiltonian can be achieved by tuning the Rabi frequency $\Omega_I$. We enforce Eq. \ref{eq:condition_for_generalized_RK} because the flipping amplitudes $t'$ and $t''$ that appear in the perturbative expansion of $\hat{H}_{Ry}$ are equal to each other. This makes it necessary that $V'$ be the same as the geoemetric mean of $V''/\lambda$ and $V''\lambda$, which is $V''$. We can ensure that Eq. \ref{eq:condition_for_generalized_RK} holds by tuning the geometry.

\section{S2: Quadrupole moment generation in CSSE of NH-QDM}

In this section, we illustrate that the configuration space skin effect in NH-QDM can be viewed as the generation of a quadrupole moment in the system. We first express the RK Hamiltonian in terms of hardcore bosonic operators $b,b^\dagger$. We take the hardcore bosons to live on the links of the square lattice. We will denote the real lattice by $\Lambda$, and the dual lattice by $\tilde{\Lambda}$. We can rewrite the kinetic term of the RK model as:
\begin{equation}
\begin{split}
      \ket{\hplaquette} \bra{\vplaquette} &=  b^{\dagger}_{\boldsymbol{R}- \hat{e}_1/2}  b^{\dagger}_{\boldsymbol{R}+ \hat{e}_1/2}  b_{\boldsymbol{R}- \hat{e}_2/2} b_{\boldsymbol{R}+ \hat{e}_2/2}, \\
      \ket{\vplaquette} \bra{\hplaquette}&= b^{\dagger}_{\boldsymbol{R}-\hat{e}_2/2}  b^{\dagger}_{\boldsymbol{R}+ \hat{e}_2/2}  b_{\boldsymbol{R}- \hat{e}_1/2} b_{\boldsymbol{R}+ \hat{e}_1/2},
\end{split}
\end{equation}
where $\hat{e}_1 \equiv \hat{x}$ and $\hat{e}_2 \equiv \hat{y}$, and $\boldsymbol{R}$ is the position of a site in the dual lattice $\Lambda$. The dimer-dimer repulsion term can be expressed as:
\begin{equation}
    \ket{\hplaquette} \bra{\hplaquette} + \ket{\vplaquette} \bra{\vplaquette}= \sum_{i=1}^2 b^{\dagger}_{\boldsymbol{R}-\hat{e}_i/2}  b_{\boldsymbol{R}- \hat{e}_i/2} 
    b^{\dagger}_{\boldsymbol{R}+ \hat{e}_i/2}  b_{\boldsymbol{R}+ \hat{e}_i/2}.
\end{equation}
The non-Hermitian RK Hamiltonian looks like the following:
\begin{equation}
\label{eq:bosonic_RK}
\begin{split}
    \hat{H} &= -t\sum_{\boldsymbol{R}} \left(e^{-\gamma}\,b^{\dagger}_{\boldsymbol{R}- \hat{e}_1/2}  b^{\dagger}_{\boldsymbol{R}+ \hat{e}_1/2}  b_{\boldsymbol{R}- \hat{e}_2/2} b_{\boldsymbol{R}+ \hat{e}_2/2} + e^{\gamma}\, b^{\dagger}_{\boldsymbol{R}-\hat{e}_2/2}  b^{\dagger}_{\boldsymbol{R}+ \hat{e}_2/2}  b_{\boldsymbol{R}- \hat{e}_1/2} b_{\boldsymbol{R}+ \hat{e}_1/2}\right)\\
    &\quad + V\sum_{\boldsymbol{R}} \sum_{i=1}^2 b^{\dagger}_{\boldsymbol{R}-\hat{e}_i/2}  b_{\boldsymbol{R}- \hat{e}_i/2} 
    b^{\dagger}_{\boldsymbol{R}+ \hat{e}_i/2}  b_{\boldsymbol{R}+ \hat{e}_i/2}.
\end{split}
\end{equation}
The hardcore bosonic operators satisfy the following commutation relation:
\begin{equation}
    [b,b^\dagger] = 1-2b^\dagger b.
\end{equation}
Note that the Hamiltonian in Eq. \ref{eq:bosonic_RK} conserves both the charge $P^{(0)} = \sum_{\boldsymbol{r}} b^{\dagger}_{\boldsymbol{r}} b_{\boldsymbol{r}}$ and the dipole moment $P^{(1)} = \sum_{\boldsymbol{r}} \boldsymbol{r} \,b^{\dagger}_{\boldsymbol{r}} b_{\boldsymbol{r}}$, where $\boldsymbol{r}$ runs over the links of $\Lambda$. The Hamiltonian in Eq. \ref{eq:bosonic_RK} can be turned symmetric by making a similarity transformation given by:
\begin{equation}
\label{eq:Qxx-Qyy}
    S = e^{\gamma \left(Q_{xx} - Q_{yy}\right)}
\end{equation}
where $Q_{xx}$ and $Q_{yy}$ are defined as:
\begin{equation}
    Q_{xx}  = \sum_{\boldsymbol{r}} r^2_x \,b^\dagger_{\boldsymbol{r}} b_{\boldsymbol{r}}, \qquad  Q_{yy}  = \sum_{\boldsymbol{r}} r^2_y \,b^\dagger_{\boldsymbol{r}} b_{\boldsymbol{r}}.
\end{equation}
To see that the similarity transformation in Eq. \ref{eq:Qxx-Qyy} works, let us explicitly act it on the bosonic operators. The action of $S$ on $b$ and $b^\dagger$ (using the fact that $b$ and $b^\dagger$ are hardcore bosonic operators) is given by:
\begin{equation}
    \begin{split}
        S \,b_{\boldsymbol{r}}\, S^{-1} &= e^{\gamma \left(Q_{xx}(\boldsymbol{r}) - Q_{yy}(\boldsymbol{r})\right)}\, b_{\boldsymbol{r}}\, e^{-\gamma \left(Q_{xx}(\boldsymbol{r}) - Q_{yy}(\boldsymbol{r})\right)} = b_{\boldsymbol{r}}\, e^{-\gamma \left(Q_{xx}(\boldsymbol{r}) - Q_{yy}(\boldsymbol{r})\right)} = e^{-\gamma(r_x^2 - r^2_y)} b_{\boldsymbol{r}}, \\
       S \,b^\dagger_{\boldsymbol{r}}\, S^{-1} &= e^{\gamma \left(Q_{xx}(\boldsymbol{r}) - Q_{yy}(\boldsymbol{r})\right)}\, b^\dagger_{\boldsymbol{r}}\, e^{-\gamma \left(Q_{xx}(\boldsymbol{r}) - Q_{yy}(\boldsymbol{r})\right)} = e^{\gamma \left(Q_{xx}(\boldsymbol{r}) - Q_{yy}(\boldsymbol{r})\right)}b^\dagger_{\boldsymbol{r}} = e^{\gamma(r^2_x - r^2_y)} b^\dagger_{\boldsymbol{r}}.
    \end{split}
\end{equation}
The terms in $\hat{T}$ transform as:
\begin{equation}
    \label{eq:kinetic_term_similarity_transform} 
    \begin{split}
    S\, b^{\dagger}_{\boldsymbol{R}- \hat{e}_1/2}  b^{\dagger}_{\boldsymbol{R}+ \hat{e}_1/2}  b_{\boldsymbol{R}- \hat{e}_2/2} b_{\boldsymbol{R}+ \hat{e}_2/2}\, S^{-1} = e^{F}\, b^{\dagger}_{\boldsymbol{R}- \hat{e}_1/2}  b^{\dagger}_{\boldsymbol{R}+ \hat{e}_1/2}  b_{\boldsymbol{R}- \hat{e}_2/2} b_{\boldsymbol{R}+ \hat{e}_2/2},
    \end{split}
\end{equation}
where $F$ is given by:
\begin{equation}
\begin{split}
    F &= \gamma\left((R_x+1/2)^2 - (R_y)^2 + (R_x-1/2)^2 - (R_y)^2 - (R_x)^2 + (R_y+1/2)^2 - (R_x)^2 + (R_y-1/2)^2\right)\\
    &= \gamma \left(2R^2_x - 2R^2_y + \frac{1}{2} - 2R^2_x + 2R^2_{y} + \frac{1}{2}\right) \\
    &= \gamma.
\end{split}
\end{equation}
Similarly, 
\begin{equation}
    S\, b^{\dagger}_{\boldsymbol{R}- \hat{e}_2/2}  b^{\dagger}_{\boldsymbol{R}+ \hat{e}_2/2}  b_{\boldsymbol{R}- \hat{e}_1/2} b_{\boldsymbol{R}+ \hat{e}_1/2}\, S^{-1} = e^{-\gamma}\, b^{\dagger}_{\boldsymbol{R}- \hat{e}_1/2}  b^{\dagger}_{\boldsymbol{R}+ \hat{e}_1/2}  b_{\boldsymbol{R}- \hat{e}_2/2} b_{\boldsymbol{R}+ \hat{e}_2/2}.
\end{equation}
Therefore, the kinetic term $\hat{T}$ transforms as:
\begin{equation}
    S \hat{T} S^{-1} = \sum_{\boldsymbol{R}} b^{\dagger}_{\boldsymbol{R}- \hat{e}_1/2}  b^{\dagger}_{\boldsymbol{R}+ \hat{e}_1/2}  b_{\boldsymbol{R}- \hat{e}_2/2} b_{\boldsymbol{R}+ \hat{e}_2/2} + b^{\dagger}_{\boldsymbol{R}- \hat{e}_2/2}  b^{\dagger}_{\boldsymbol{R}+ \hat{e}_2/2}  b_{\boldsymbol{R}- \hat{e}_1/2} b_{\boldsymbol{R}+ \hat{e}_1/2}.
\end{equation}
The similarity transformation leaves the dimer-dimer repulsion term unchanged. Therefore, the configuration space skin effect in the NHRK model is in agreement with the results derived in Ref. \cite{jacopo1} in the context of multipole-conserving non-Hermitian many-body dynamics.

\bibliography{references} 